\documentclass[12pt]{article}
\usepackage{setspace}
\usepackage{fancyhdr}
\usepackage{booktabs}
\usepackage[margin=1in, headheight=15pt]{geometry}
\usepackage{lipsum}
\usepackage{ragged2e}

\usepackage{graphics}
\usepackage{tikz}
\usetikzlibrary{calc}
\usetikzlibrary{decorations.pathreplacing}
\usetikzlibrary{arrows.meta}
\usepackage{amsmath,amsfonts,amsthm}
\usepackage{mathrsfs}
\usepackage{cancel}
\usepackage[normalem]{ulem}
\usepackage{siunitx}
\usepackage{float}
\usepackage{enumitem}
\usepackage{colortbl}
\usepackage{subcaption} 
\usepackage[font=footnotesize,labelfont=bf]{caption}
\usepackage{tabularx}
\newcolumntype{L}[2]{>{\hsize=#1\hsize\columncolor{#2}\raggedright\arraybackslash}X}%
\newcolumntype{R}[2]{>{\hsize=#1\hsize\columncolor{#2}\raggedleft\arraybackslash}X}%
\newcolumntype{C}[2]{>{\hsize=#1\hsize\columncolor{#2}\centering\arraybackslash}X}%

\usepackage{color}
\definecolor{mygreen}{rgb}{0,0.6,0}
\definecolor{mygray}{rgb}{0.5,0.5,0.5}
\definecolor{mymauve}{rgb}{0.58,0,0.82}
\definecolor{mylightgray}{gray}{.9}
\definecolor{myblue}{rgb}{.28,.24,.55}
\definecolor{mylightblue}{rgb}{0.74, 0.83, 0.9}
\definecolor{mybrightred}{rgb}{1,.13,.32}
\definecolor{mypink}{rgb}{0.96, 0.76, 0.76}
\definecolor{mylightgreen}{rgb}{0.66, 0.89, 0.63}
\usepackage{colorspace}
\definespotcolor{clear}{rgb}{0,0,0,0}

\usepackage{natbib}
\setcitestyle{aysep={}}
\newcommand\posscite[1]{\citeauthor{#1}\textcolor{darkgray}{'s} (\citeyear{#1})}
\PassOptionsToPackage{hyphens}{url}
\usepackage{hyperref}
\hypersetup{colorlinks=true, citecolor=darkgray, linkcolor=darkgray, urlcolor=darkgray}
\makeatletter
\let\@fnsymbol\@arabic
\makeatother

\usepackage{lmodern}
\usepackage{listings}
\usepackage[textsize=tiny, backgroundcolor=mypink, linecolor=red]{todonotes}

\newcommand{\mytitle}{Spread of Tweets in Climate Discussions}
\newcommand{\shorttitle}{Spread of Tweets in Climate Discussions}
\author{Yan Xia\thanks{Department of Computer Science, Aalto University}, Ted Hsuan Yun Chen\footnotemark[1]\kern 4pt \thanks{Faculty of Social Sciences, University of Helsinki} \kern -9pt, \kern 3pt and Mikko Kivel\"a\footnotemark[1]}
\newcommand{\surname}{Xia, Chen, and Kivel\"a}

\title{\mytitle}
\date{}

\begin{document}
\pagenumbering{roman}
\singlespacing
\maketitle
\thispagestyle{empty}
\begin{abstract}
\noindent Characterising the spreading of ideas within echo chambers is essential for understanding polarisation. In this paper, we explore the characteristics of popular and viral content in climate change discussions on Twitter around the 2019 announcement of the Nobel Peace Prize, where we find the retweet network of users to be polarised into two well-separated groups of activists and sceptics. Operationalising popularity as the number of retweets and virality as the spreading probability inferred using an independent cascade model, we find that the viral themes echo and differ from the popular themes in interesting ways. Most importantly, we find that the most viral themes in the two groups reflect different types of bonds that tie the community together, yet both function to enhance ingroup connections while repulsing outgroup engagement. With this, our study sheds light, from an information spreading perspective, on the formation and upkeep of echo chambers in climate discussions. 


\end{abstract}

\newpage
\tableofcontents

\newpage
\onehalfspacing
\pagenumbering{arabic}
\setcounter{page}{1}

\section{Introduction}
Climate discussions on social media platforms are often structured into segregated echo chambers of activists and sceptics \citep{williams2015network}. At best, these echo chambers simply demonstrate that different opinions on climate politics exist; worse, they cause vicious cycles of increasing divisiveness \citep{asikainen2020cumulative, baumann2020modeling}. Better knowledge of the generative features of polarised echo chambers can facilitate the design of mitigation institutions, especially on social media platforms where the information individuals are exposed to can be algorithmically manipulated \citep{musco2018minimizing, nelimarkka2018social}. However, while prior work demonstrates the existence of these echo chambers, we lack a nuanced understanding of the social processes that generate them. Prior studies point to attitude-based homophily whereby individuals limit their interactions to others with whom they share similar attitudes \citep{williams2015network} or entrenched group cleavages resulting from issue alignment \citep{chen2020polarization} as important features of climate discussion echo chambers, but these mechanisms alone cannot explain the complexity of the observed discussion networks. 

On social media platforms where information sharing serves as a major channel of communication, a better understanding of the discussion dynamics could be achieved by observing how different types of information spread through the network, accordingly consolidating or attenuating the echo chambers. Thus, in order to better understand the sustained existence of echo chambers in climate discussions on social media platforms, we conduct a study on Twitter that examines climate discussion dynamics through the lens of tweet spreading, particularly what kind of content is most likely to be shared within and across different groups of users. We focus our examination on the core of the echo chambers which comprises the most popular tweets and users, drawing on research showing that climate politics polarises following elite behaviour \citep{birch2020political}.

More specifically, given a division of the users into climate activists and climate sceptics, we examine what tweets spread within the echo chambers, what cross the boundary, and what characteristics of a tweet lead to more viral spreading on the user network. On the last point, in contrast to previous studies that quantified the virality of an item as the times it is shared \citep[e.g.][]{berger2012makes, hansen2011good}, we measure virality while controlling for the underlying complex social network. This allows us to account for the number of times and the context under which a tweet has been seen.

We conduct our study on the climate discussions during the announcement of the 2019 Nobel Peace Prize. The event triggered extensive discussion on climate politics because Greta Thunberg, the climate activist who mobilised demonstrations around the world, was recognised in the media as a likely winner \citep{adam2019why, carrington2019greta}. The fact that Thunberg did not win spurred a host of new discussion points among climate change sceptics. For this reason, the event provides a suitable opportunity for our study, as it injected new information into extant discussion networks. During that period, multiple opinions, attitudes, and emotions from different user groups collided and merged in online social networks, thereby providing a rich context for studying the spreading dynamics of different types of information in climate discussions. 

Our study makes a number of contributions. First, it adds to the body of work describing the content of climate discussions on Twitter \citep[e.g.][]{cody2015climate, dahal2019topic, jang2015polarized, kirilenko2014public}. Second, it provides a finer picture of climate discussion dynamics through information spreading, furthering our understanding of echo chambers in online discussions \citep{barbera2015tweeting}, especially on climate politics \citep{williams2015network}. Here, it also links more broadly to the literature on the relationship between filter bubbles and political polarisation, which is the concept that the  feedback between behavioural selective exposure and algorithmically-learned filtering of information will lead to societal polarisation \citep[e.g.][]{dahlgren2021critical, pariser2011filter}. While our study does not consider the algorithmic side of these filter bubbles, we find strong evidence of selective engagement with ingroup members, and to the extent that outgroup information crosses group cleavages, it potentially contributes to a greater degree of inter-group hostility. Finally, our study complements previous work on the virality of online content \citep{berger2012makes, hansen2011good, stieglitz2013emotions} in adopting a network-aware measure of virality. 

The remainder of this paper proceeds in two steps. First, we explore the characteristics of popular tweets in each of the echo chambers using an iterative process. After collecting all tweets containing “climate” during the announcement period and identifying the activists and sceptics involved, we examine the characteristics of the discussion within the two groups using an initial reading of the most popular tweets. Second, using statistical methods, we infer the virality of tweets from our sample and examine which of the previously identified characteristics most strongly predict likeliness to spread. We conclude by discussing the implications of our findings. Most importantly, our results show that the most virality-predicting features of a tweet are also ones that enhance ingroup ties while repulsing outgroup engagement, thus revealing the potential role of tweet spreading in exacerbating the polarisation on climate Twitter.

\section{Exploring Climate Twitter}
While there have been a number of studies that describe climate discussions on Twitter \citep[e.g.][]{cody2015climate, dahal2019topic, jang2015polarized, kirilenko2014public}, many of them are more than a few years old. The rapid rate of change of online behaviour means that many of these findings have decreased temporal validity \citep{munger2019limited}. We therefore begin our study with an exploratory approach to understanding the present state of climate discussions. Our primary goal in this exercise is to identify popular themes in current climate discussions such that we can use them to study tweet spreading dynamics. Our approach is an iterative theory-building process, where prior literature sets our expectations for a first-cut exploratory reading of our data, which in turn informs our expectations about tweet virality and our codebook development.

\subsection{Literature review}
We begin our study with a literature review that informs our data exploration, focusing on identifying popular themes in climate discussion, how they differ between activist and sceptic groups, and more generally the determinants of information virality on online social network platforms. Earlier work on Twitter-specific climate discussions tends to be more descriptive, with studies laying the groundwork by describing the geospatial distribution of tweets and variations in their content \citep[e.g.][]{kirilenko2014public}. \citet{jang2015polarized}, for example, showed that among four English-speaking countries (U.S., U.K., Canada, and Australia), there is considerable variation in how people tweet about climate change. Most relevant to polarisation, they found that, within the U.S., Republican- and Democrat-leaning states exhibit subnational differences, with the former more likely to engage with the “hoax” frame. These results comport with studies showing that the climate issue is generally subject to partisan sorting \citep[e.g.][]{mccright2011politicization}. Related to this are observed discrepancies in references to “science”. In their review of climate-related tweets, \citet{cody2015climate} found that sceptics tend to use phrases that are less commonly used in conjunction with references to science. On the other hand, \citet{pearce2014climate}, possibly due to the more science-focused nature of their case study, found that in discussions of IPCC reports, science was highly politicised. Some studies focused explicitly on affective polarisation \citep[e.g.][]{tyagi2020affective}, finding large variation between activist and sceptic groups in expressing negativity toward their respective outgroups. 

Observed differences between groups are telling but, with the exception of \posscite{williams2015network} work on homophily, these studies do not explicitly address the formation of echo chambers. Drawing on the elite-led polarisation literature \citep{birch2020political}, we propose that communities resembling echo chambers form when tweets (more generally content) from core members of the group spread to the periphery. This kind of spreading mechanism, coupled with homophily in the sharing network, which reduces cross-group ties, should result in the kind of echo chamber structure that has been observed in climate discussions. The implication here is that without the spread of viral tweets from the core, the echo chambers will dissipate, making the study of what gets shared an important area of climate polarisation research.

Previous work in this area, beyond climate discussions, has shown that in online social networks, the spreading of content is strictly connected to the characteristics of the content itself \citep{guerini2011exploring, jenders2013analyzing}, including types of content \citep{nagarajan2010qualitative}, URLs, hashtags \citep{suh2010want}, and emotions \citep{berger2012makes, hansen2011good, stieglitz2013emotions}. Specifically, the textual features of content largely predict its spreading on Digg \citep[a news based internet sharing platform,][]{guerini2011exploring}, and the type of a tweet (e.g. calling for action, information sharing) plays a vital role in shaping its retweet network on Twitter \citep{nagarajan2010qualitative}. Meanwhile, others find URLs and hashtags to be the most important content features correlated with retweetability on Twitter \citep{suh2010want}. Hashtags have also been identified as a way individuals engage with a broader imagined community on social media platforms \citep{hanteer2019innovative}. Various emotions (i.e. positive/negative sentiments, and more specifically awe, anger, anxiety) in the content also positively predict sharing behavior of users on both New York Times \citep{berger2012makes} and Twitter \citep{hansen2011good, stieglitz2013emotions}. Other than the content features, author features including account age, follower count, and followee count also affect how likely content is shared on Twitter \citep{jenders2013analyzing, suh2010want}. Finally, information diffusion on social networks is influenced by their underlying network structures \citep{weng2013virality}. 

\subsection{Data}
To further explore the concepts identified in our literature review, we collected Twitter data on climate discussions. As noted, we focused on the climate discussion surrounding the 2019 Nobel Peace Prize announcements because Greta Thunberg featured prominently as a likely winner. Using the Twitter streaming application programming interface (API) and a list of relevant keywords, we collected every tweet related to Greta Thunberg, climate politics, and the Nobel Peace Prize from Oct. 10 to Oct. 22 2019, which covers a short interval before and after the announcement of the 2019 Nobel Peace Prize on Oct. 11 2019. A total of 5,422,617 tweet records were collected, including records of both original tweets and retweets, and each record contains information of the text, author and status of a tweet (e.g. whether it is an original tweet or a retweet). A total of 2,011,410 users were involved in the data set.

Since our study aims to explore discussion dynamics with respect to climate change, we focused our examination on tweets that include the substring “climate” (case-insensitive) in its text. We further restricted our examination to tweets written in English, and excluded replies to other tweets. Based on trends identified from prior work \citep{jang2015polarized}, we also selected and focused on the subset of users who have either posted or retweeted any tweet with hashtags that included “climate” in combination with either “crisis” or “hoax” (case-insensitive). Using this set of 317,243 tweet records with this set of 24,770 users, we built a retweet network of users, with each user corresponding to a single node, and a single link connecting two nodes if there exists any retweet record going between the corresponding users (i.e. an undirected link exists between node A and node B if either user A retweeted any tweet posted by user B at least once, or user B retweeted any tweet posted by user A at least once).
Following prior work in this area \citep[e.g.][]{chen2020polarization, garimella2018quantifying}, we focused on only the largest connected component of the retweet network, which consists of 20,628 nodes (i.e. users) and 73,381 links, as visually shown in Figure 1; the disconnected components were excluded because there was not sufficient information for inferring their stances on the topic. Colours of the user nodes indicate climate activism or scepticism, classified using a network partitioning method. Following prior work \citep[e.g.][]{garimella2018quantifying}, we used the METIS partitioning algorithm which finds the two groups with the lowest intergroup connections \citep{karypis1998fast}. After the partitioning, the green bubble consists of 14,812 users classified as climate activists, and the orange bubble consists of 5,816 users classified as climate sceptics. A total of only 223 links going between the two bubbles indicates that our procedure yielded two echo chambers of climate activists and sceptics.

\begin{figure}[!htb]
  \centering
  \includegraphics[width=.85\columnwidth]{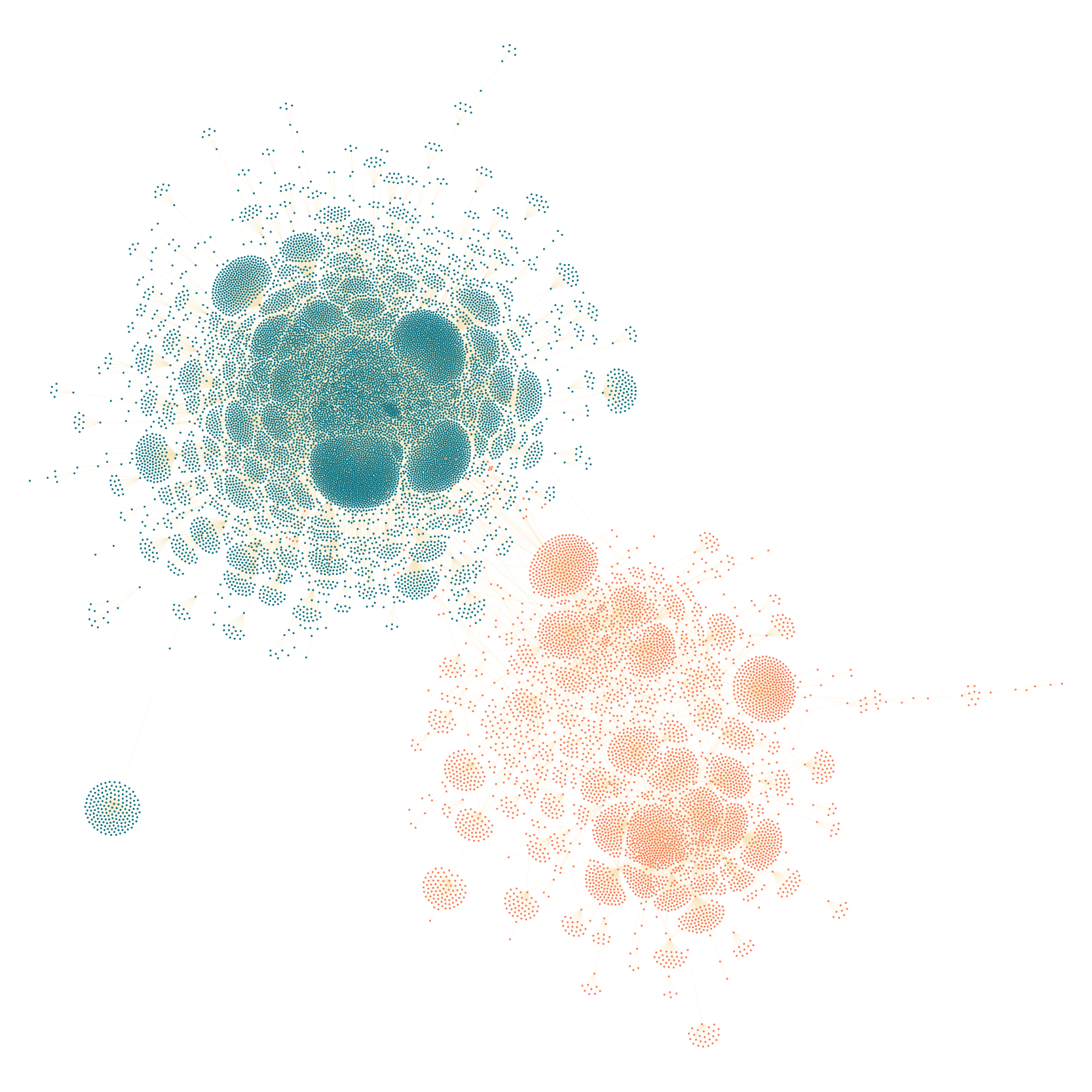}
  \caption{The partitioned retweet network of selected users and tweets, where the green bubble consists of 14,812 users classified as climate activists, and the orange bubble consists of 5,816 users classified as climate sceptics.}
  \label{fig:network}
\end{figure}

\subsection{Within-group spreading tweets and popular themes}
\label{sec:labeling}
From the data described above, we prepared a subset of tweets for study. Because we are interested in understanding how tweets from popular users spread, we focused on the top 500 most retweeted items in each group. These have in-group retweet counts ranging from 66 to 1,762 in the activist group, and from 26 to 520 in the sceptic group.

\begin{figure}
  \centering
  \includegraphics[width=\columnwidth]{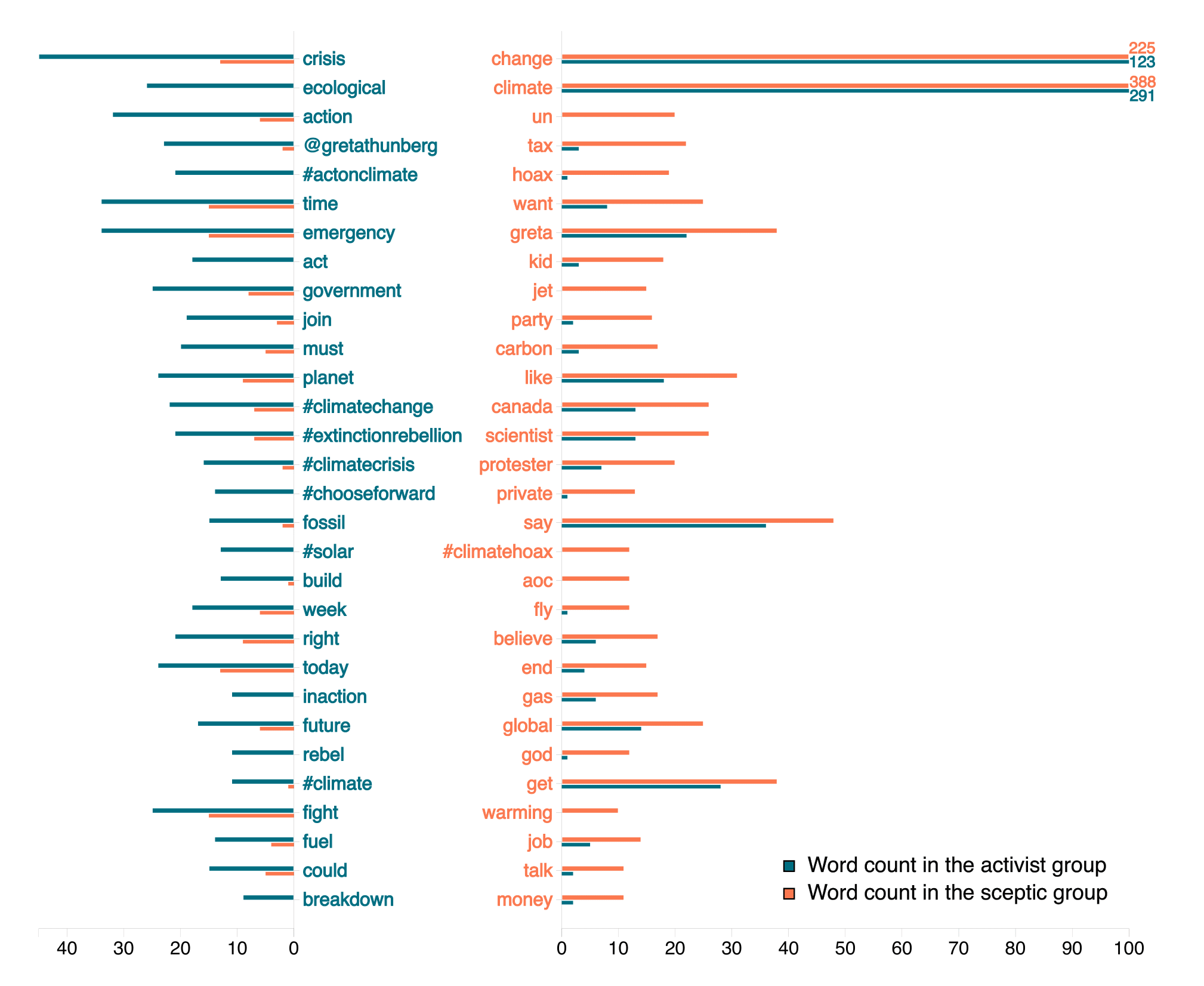}
  \caption{Most characteristic words among the top 500 retweeted tweets in respectively the activist group (green) and the skeptic group (orange), ranked by the difference between word counts in the two groups. The bars of ``change" and ``climate" with exceptionally large word counts are truncated for visualisation purposes, but the actual word counts are attached in number to the right of the bars. }
  \label{fig:popwords}
\end{figure}

We begin our examination by identifying the most commonly occurring words in each group (counted maximum once per tweet), adjusting for how often they appear in the other group. These words are presented in \autoref{fig:popwords}. There seems to be a clear theme of taking action among the activist discussions, with ``crisis'', ``action'', ``\#actonclimate'' and ``act'' among the ten most common words. In this group, Greta Thunberg's Twitter account is frequently mentioned. In the sceptic group, there is frequent mention of actors and entities from the other side (e.g. ``greta'', ``aoc'', ``protester'', ``un''). Our later reading shows that these mentions result from outgroup attacks. The United Nations, for example, is often discussed from a negative, anti-internationalist perspective. The sceptics also commonly used ``jet'', ``private'', and ``fly'' to support their argument that activists are hypocritical about environmental protection. It is also interesting that the words ``climate'' and ``change'' appear much more in the sceptic group than in the activist group, because activists tend to use ``climate crisis'' and ``\#climatechange'' instead of ``climate change''.

Next, we conducted an exploratory reading of the tweet texts to identify commonly occurring themes and other characteristics using approximately half the tweets in each group. This procedure was informed by our literature review, and as expected, we encountered many of the same themes as those outlined in prior research. However, our reading did not always comport with the existing literature. For example, whereas \citet{cody2015climate} showed indirect evidence for discrepancies between activists and sceptics in references to ``science'', we found that both groups invoked the concept of science at a similar rate. This is surprising, as we initially expected to see more activists invoking scientific evidence that the climate crisis actually exists. Instead the sceptic group also made frequent references to scientific evidence they claimed to disprove anthropogenic climate change. This observed discrepancy to prior work offers evidence of temporal shifts in the climate discussion, justifying our initially exploratory approach.

In general, the most significant outcome of our reading is the clear difference between the discussion styles and content in the two groups. The dominating sentiment in the sceptic group is negative and even aggressive, while the sentiment in the activist group is more positive. In terms of content, the most popular tweets among the activists share the theme of ``action''. They either celebrated actions taken to tackle climate change, especially the pro-climate movements, or called for further action to address the climate crisis, potentially proposing concrete solutions. For example, the global climate strike led by Greta Thunberg was a popular topic among the activists. A large number of tweets are images or videos that demonstrate the strength of the movement or praise certain individuals who participated in the movement, such as the protesters (especially celebrities) who got arrested. The pro-climate speeches by Greta Thunberg and Alexandria Ocasio-Cortez, the U.S. senator who spoke out in favour of strong climate policies, were also widely quoted and commended. Additionally, another set of tweets proposed solutions to address climate change and calls for action to implement them.

Among the sceptics, the most popular tweets are dominated by the theme of ``attack". Most of them are directed sharply at climate change supporters, either making fun of them, or accusing them of being hypocritical by engaging in environmentally-harmful practices. For example, we saw a large number of attacks and mocking toward climate activists including Greta Thunberg, Alexandria Ocasio-Cortez, and Jane Fonda. Other pro-climate celebrities were accused for flying in private jets. Aside from this, there are also recurring patterns that claim climate policy to be a front for other agendas (e.g. money, control), speak negatively of international organisations (e.g. the UN), or invoke ``science" or ``scientist" to back up the argument. With respect to the style of language used, there is a dominating theme of mocking tones, potentially also paired with the use of uncivil wording, emojis, or exclamation marks.

Based on this exploratory reading process, we identified a list of tweet features we deemed to be plausibly relevant to a tweet's virality. \autoref{tab:labels} contains these features along with the coding rule we developed for identifying them. 

\begin{table}[!htb]
    \centering \footnotesize \renewcommand{\arraystretch}{1.1}
    \begin{tabular}{l l} \toprule
         Feature & Coding Rule \\ \hline
         \multicolumn{2}{l}{\textbf{Universal (Style)}} \\
         Mocking & Does the tweet make fun of an entity at its expense? (yes/no)\\
         Incivility & Does the tweet contain uncivil language? (yes/no) \\ \\
         
         \multicolumn{2}{l}{\textbf{Universal (Content)}} \\
         Call to Action & Does the tweet call on others to behave in a certain way? (yes/no) \\
         Ingroup Praise & Does the tweet speak of the ingroup in a positive manner? (yes/no) \\
         Outgroup Criticism & Does the tweet speak of the outgroup in a negative manner? (yes/no) \\
         Science & Does the tweet invoke ``science'' or ``facts'' as support? (yes/no) \\ 
         Hashtags* & Machine extracted count of hashtags used in the tweet. \\ 
         Mentions* & Machine extracted count of users mentioned in the tweet. \\ \\
         
         \multicolumn{2}{l}{\textbf{Activists}} \\
         Solutions & Does the tweet present solutions to addressing climate change? (yes/no) \\
         Movement & Does the tweet emphasise the strength of the pro-climate movement? (yes/no) \\ \\
         
         \multicolumn{2}{l}{\textbf{Sceptics}} \\
         Anti-international & Does the tweet speak negatively of international organisations? (yes/no) \\
         Hypocrisy & Does the tweet claim that supporters are hypocritical or inconsistent? (yes/no) \\
         Conspiracy & Does the tweet claim that climate policy is a front for other agendas? (yes/no) \\
         \bottomrule
         \multicolumn{2}{l}{* indicates features directly extracted from the data.}
    \end{tabular}
    \caption{Tweet features for coding.}
    \label{tab:labels}
\end{table}

To conduct more systematic analysis on characteristics of the echo chamber cores, we prepared our final data set by further filtering out tweets by users who authored fewer than three tweets. In this way, we are defining the core of the echo chambers as individuals who are are consistently responsible for the popular tweets in either group. This filtering step also aids the later virality analysis because it allows us to compare the importance of features while holding user characteristics constant. For each of these tweets, we did a second reading, labelling the based on whether they contain each of our previously identified features following the coding rules outlined in \autoref{tab:labels}. When encountering external links, we first decide whether the tweet endorses or criticises the external content. If it is endorsed, we take the external content as an extension of the tweet when labelling. We decided to adopt a manual coding approach for labelling the features of each tweet, with the observation that most features we had selected to code involved a nuanced understanding of the tweet content. As much as lexicon-based and machine learning methods had been employed in previous studies to automatically detect simpler features of text, including positivity and negativity \citep{berger2012makes, hansen2011good, stieglitz2013emotions}, those methods are much less effective in recognising features that involve richer emotions and contextual knowledge. Other studies that included similarly nuanced features, such as incivility, also relied entirely or in part on human-coding \citep{berger2012makes, muddiman2019re}. 

\begin{figure*}
  \centering
  \includegraphics[width=.85\columnwidth]{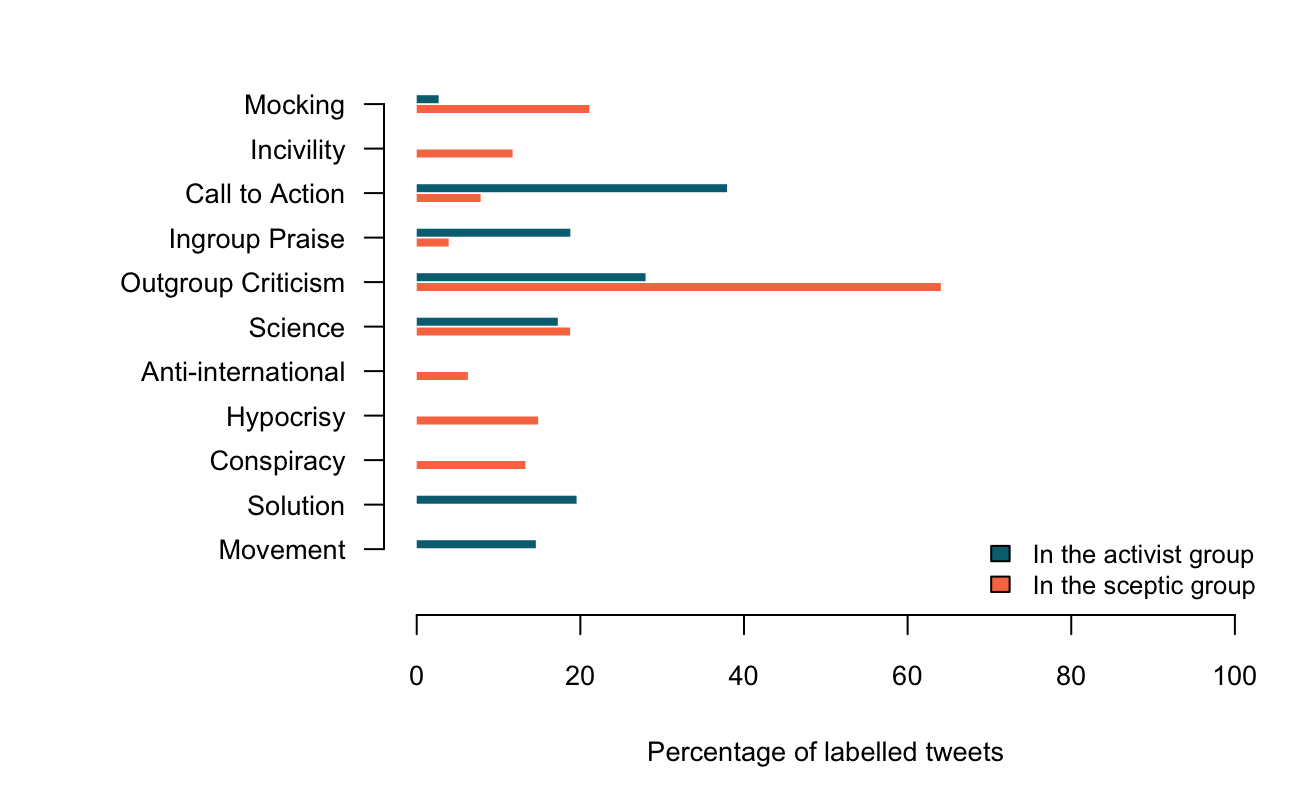}
  \caption{Percentage of labelled tweets in each group with respect to each variable.}
  \label{fig:var}
\end{figure*}

In this second step, tweets were labelled independently by all three authors, which allowed us to adjudicate disagreements using a vote. Overall, we saw a moderate level of agreement prior to voting, with the inter-rater reliability score Krippendorff's alpha being 0.63, and all three authors reaching a consensus on 86.2\% of the entries. Finally, in addition to these manually labelled features, we used computer text processing to directly extract the number of hashtags used and the number of other users mentioned in the tweet. \autoref{fig:var} shows the results of our labelling process. Our qualitative findings based on our exploratory reading, on the parts that are relying on popularity of content, are confirmed by these quantitative results.

\subsection{Cross-group spreading tweets}
\begin{figure*}
  \centering
  \includegraphics[width=0.75\columnwidth]{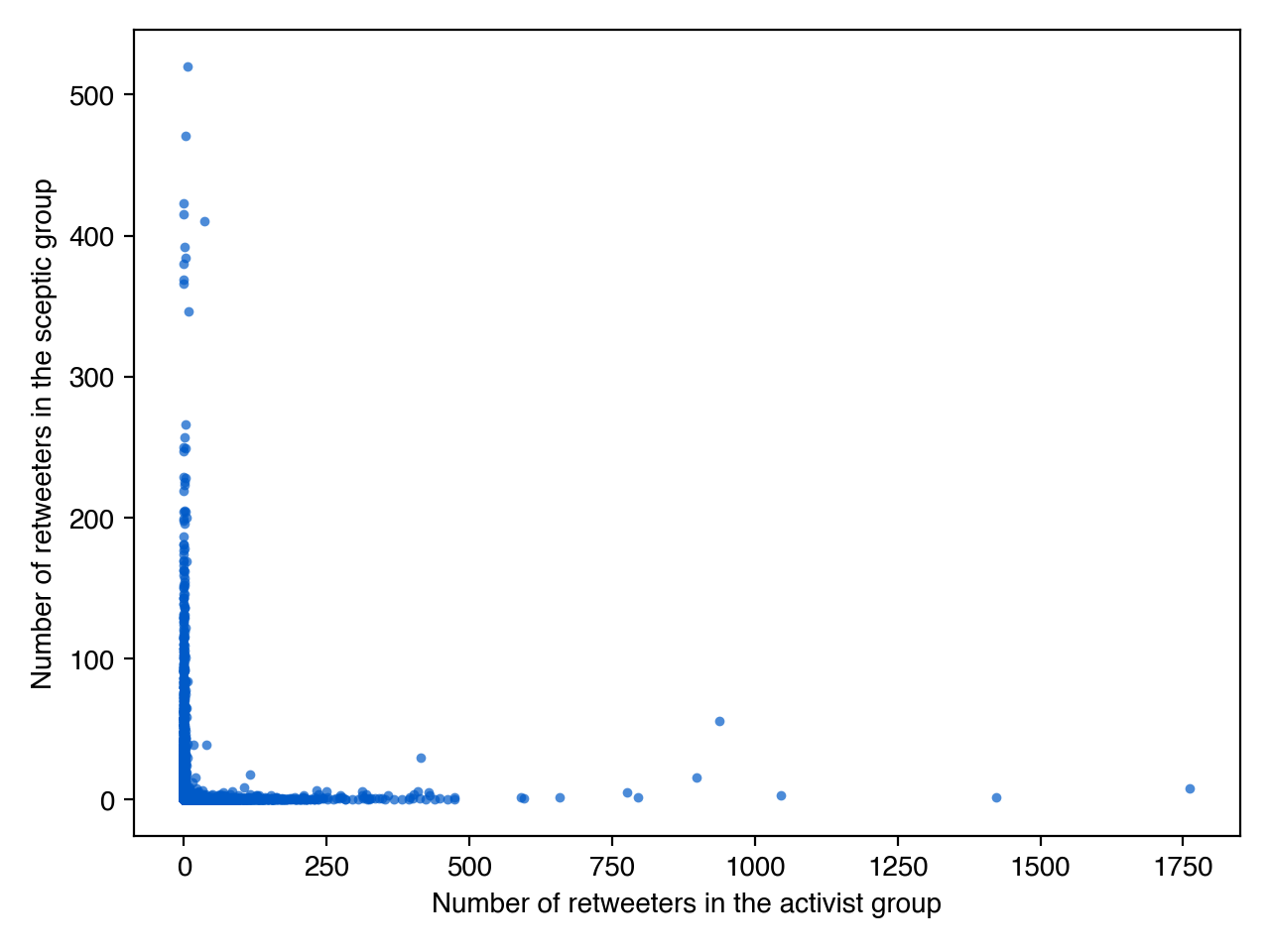}
  \caption{Scatter plot of tweets with respect to number of retweeters in each group.}
  \label{fig:rt}
\end{figure*}

The largely isolated bubble structures apparent in \autoref{fig:network} already suggests that cross-group spreading is rare. We examined this more deeply by checking the number of tweets spreading in both groups. The results, shown in \autoref{fig:rt}, which is a scatter plot of tweets with one axis being the number of retweeters in the activist group and the other axis being the number of retweeters in the sceptic group, indicate that most tweets in our data set mainly spread in only one group, and only only 9 tweets get retweeted more than 10 times in both groups.

Then, we inspected the tweets that mainly spread in one group but also got a significant number of retweets from the other group to see if there exists any meaningful pattern. Among these cross-group spreading tweets, the only recurring pattern we recognised is that they documented some less successful or more extreme approaches to protests taken by the climate activists (e.g. protesters being arrested, protesters burying their heads in the sand). Further, the direction of spread of these tweets ran from the activist group to the sceptic group. This finding is tentative given the small sample of data, but we speculate that these tweets were “popular” with both groups because even without contextualisation via quote-retweeting, they allowed the respective groups to interpret it as they want. However, this cross-group information flow presumably further intensifies the conflict between activists and sceptics, instead of bridging the gap by fostering communication and mutual understanding.

\section{Viral Spreading on Climate Twitter}

Next, we study what kinds of content predict a tweet's viral spreading within the two groups. In doing so, we distinguish virality from the popularity measure in the preceding section. This is because the popularity of a tweet, as measured by the number of retweets, could be an inappropriate metric of how viral the tweet content actually is: there are a number of factors that influence how popular a tweet will become aside from its content. The tweet can be made by an authoritative person or organisation, or it can start from favourable location in the network such as close to actively retweeting users or from an account with a large number of followers (Jenders et al., 2013; Suh et al., 2010). Further, the design choices Twitter has made on how to show tweets to the users can have an effect on the popularity. In addition to these confounding factors, random fluctuations in the tweeting behavior are going to be amplified in spreading processes, because the more the tweet is retweeted the more people will see it and have possibility to retweet it further. This type of rich-get-richer phenomenon is well-known in spreading processes and can lead to situations where small changes in the virality can lead to large changes in popularity or large fluctuations in the popularity of similar content \citep{barrat2008dynamical}.

Here, we instead measure the \emph{virality} of a tweet, as proportional to the probability that an average user retweets the tweet after seeing it. With this information we can find out which type of content, as categorised by the labelling in Section~\ref{sec:labeling}, is more likely to spread within the two polarised groups.

\subsection{Inferring tweet virality}
Our goal is to infer the extent to which a tweet is retweeted for its content, apart from confounding contextual factors. These factors include user characteristics such as how authoritative the original tweeting account is and the retweeting account's retweeting tendencies. We also model how the tweet is shown to users. In the end a virality score of a tweet is computed by finding the score value that best explains the whole process of the tweet spreading in the network.

The core of our model uses two pieces of information. The first, how often a tweet is retweeted and by whom, is directly observable from our data. The second, how often a tweet is seen and by whom, requires that we define an exposure pathway network based on who follows who and how the tweets are shown to followers.
In Twitter, a tweet is shown to users that follow the original tweeter or anyone retweeting the original tweet. To model this, we create a follower network with directed links from users to their followers.
Such a follower network is not enough, however, because Twitter limits the times a user sees an original tweet, regardless of how many of the user's followees retweeted it. To gain the necessary understanding of the relationship between followee tweeting behaviours and what is displayed on the follower's timeline, we tested Twitter's algorithm for displaying tweets and retweets. \autoref{fig:exposure} contains our findings, which can be summarised in two rules:

\begin{figure}[!htb]
    \centering
    \begin{tikzpicture}
    \tikzstyle{origin}=[circle,draw,inner sep=3,fill=black]
    \tikzstyle{retweeter}=[circle,draw,inner sep=3,fill=lightgray]
	\tikzstyle{potential}=[circle,draw,inner sep=3,fill=white]
	
	\tikzstyle{origin_l}=[circle,draw,inner sep=2,fill=black]
    \tikzstyle{retweeter_l}=[circle,draw,inner sep=2,fill=lightgray]
	\tikzstyle{potential_l}=[circle,draw,inner sep=2,fill=white]
    
    \node[origin] (o1) at (-6, 1.5){};
    \node[potential] (p1) at (-6, -1.5){};
    \draw[-{Latex[length=1.75mm]}](o1) -- (p1);
    
    
    \node[origin] (o2) at (-3, 1.5){};
    \node[potential] (p2) at (-3, -1.5){};
    \node[retweeter] (i21) at (-2.25, 0){};
    \node[retweeter] (i22) at (-1.25, 0){};
    \draw[-{Latex[length=1.75mm]}](o2) -- (p2);
    \draw[-{Latex[length=1.75mm]}](o2) -- (i21);
    \draw[-{Latex[length=1.75mm]}](o2) -- (i22);
    \draw[-{Latex[length=1.75mm]}](i21) -- (p2);
    \draw[-{Latex[length=1.75mm]}](i22) -- (p2);
    \draw[dotted, line width= 0.65 mm](-1.95, 0) -- (-1.55, 0);
    
    \node[origin] (o3) at (1.5, 1.5){};
    \node[potential] (p3) at (1.5, -1.5){};
    \node[retweeter] (i3) at (2.25, 0){};
    \draw[-{Latex[length=1.75mm]}](o3) -- (i3);
    \draw[-{Latex[length=1.75mm]}](i3) -- (p3);
    
    \node[origin] (o4) at (5, 1.5){};
    \node[potential] (p4) at (5, -1.5){};
    \node[retweeter] (i41) at (5.75, 0){};
    \node[retweeter] (i42) at (6.75, 0){};
    \draw[-{Latex[length=1.75mm]}](o4) -- (i41);
    \draw[-{Latex[length=1.75mm]}](o4) -- (i42);
    \draw[-{Latex[length=1.75mm]}](i41) -- (p4);
    \draw[-{Latex[length=1.75mm]}](i42) -- (p4);
    \draw[dotted, line width= 0.65 mm](6.05, 0) -- (6.45, 0);
    
    \node[origin_l, label = {right}:{\tiny Original Tweeter}] at (-2.5, -2.5){};
    \node[retweeter_l, label = {right}:{\tiny Retweeter}] at (0.1, -2.5){};
    \node[potential_l, label = {right}:{\tiny Potential Retweeter}] at (1.9, -2.5){};
    \draw[-{Latex[length=1.75mm]}] (4.8,-2.5) -- (5.4,-2.5) node[right]{\tiny Exposure Pathway};
    
    \draw[] (-2.9,-2.7) -- (-2.9,-2.3) -- (8,-2.3) -- (8,-2.7) -- cycle;
    
    \draw[gray](0, -1.9) to (0, 2.1);
    \node[] at (-4.2, 2.9){\scriptsize \underline{Direct exposure pathway from origin to potential retweeter}};
    \node[] at (4.2, 2.9){\scriptsize \underline{Indirect exposure pathway from origin to potential retweeter}};
    
    \node[anchor = west] at (-8, 2.2){\scriptsize \textit{Scenarios:}};
    \node[anchor = west] at (-8, -3.1){\scriptsize \textit{Appearance:}};

    
    \end{tikzpicture}
    
    \vspace*{-8pt}

    {\includegraphics[width = 0.5\textwidth, trim = {0cm 0cm 1.5cm 0.2cm}, clip]{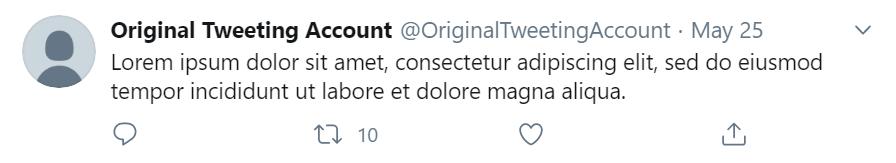}%
    \includegraphics[width = 0.5\textwidth, trim = {0cm 0cm 1.5cm 0.2cm}, clip]{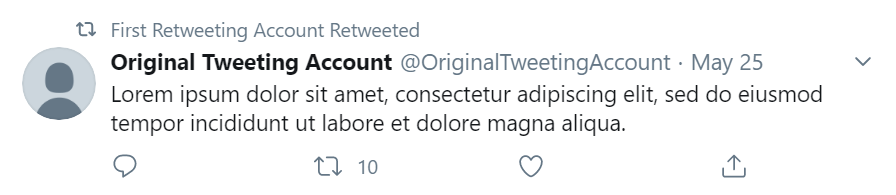}}
    \caption{Different tweeting and retweeting scenarios and what appears on the follower's timeline. In both cases, additional retweeting beyond the source closest to the original tweet is not visible on the timeline.}
    \label{fig:exposure}
\end{figure}
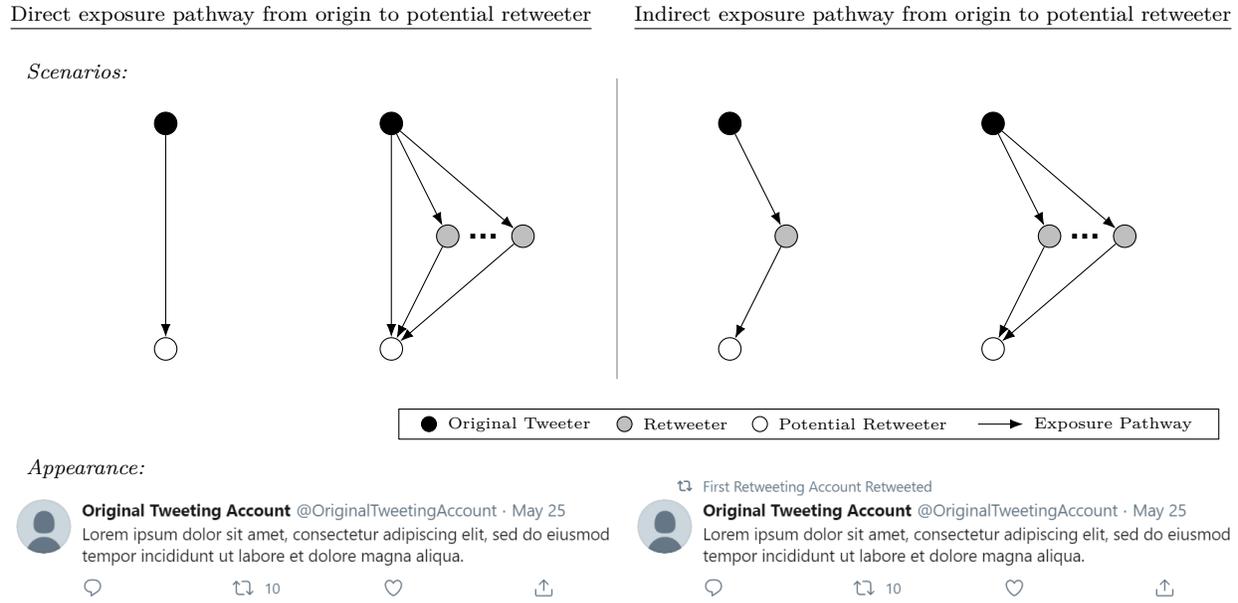

\begin{itemize}
    \item[1.] If an account makes an original tweet, followers can only see the original tweet (i.e. no retweet notifications) regardless of whether others have retweeted it. No information about retweets of the original tweet will be available on the user's timeline.
    \item[2.] If an original tweet is retweeted by one or more followees, the follower's timeline can only show the retweet notification from the first retweeting followee (and, by the first rule, only if the original tweeter is not a followee). 
\end{itemize}

A cascade, i.e., the whole process of a single tweet spreading on the follower network, is modelled similar to the independent cascade model \citep{kempe2003maximizing} and its extension \citep{barbieri2013topic}.
In our model, a user retweets a tweet they have seen with a probability that depends on the virality score of the tweet and their own general activity level.
More specifically, we first defined the activity $\alpha_u$ of user $u$ as the total number of times they have tweeted or retweeted during our data collection period.
Then, a tweet $T$ is retweeted with probability $\alpha_u r_T$,  a product of the activity of the exposed user $\alpha_u$ and the virality score of the tweet $r_T$.
The likelihood of observing a cascade in our data is computed as the joint probability of exposed users retweeting or failing to retweet independently of each other. For example, 
if in our data we observe tweet $T$ to successfully activate users $u$ and $v$, and fail to activate user $w$ in the cascade, then the likelihood of the cascade is $\mathcal{L}(r_T)=\alpha_u r_T \alpha_v  r_T (1-\alpha_w r_T)$.

Under this model, we can infer the virality of any tweet in our data set by observing the number of successful and failed activations of the tweet and the activity levels of the corresponding users. Intuitively, the more successful activations over failed activations of a tweet we observe, and the less active users the tweet manages to activate, the higher virality the tweet should have. Technically, we achieved such inference of tweet virality using a maximum likelihood method \citep{saito2008prediction}. Specifically, for every tweet $T$ in our data set, with $r_T$ as an unknown parameter, we first find the likelihood function $\mathcal{L}(r_T)$ of its cascade similar to the above example, and then find the value of $r_T$ that maximises the value of this function. This value of $r_T$ is the inferred virality score of the tweet.

The virality score of a tweet is likely not the same within and between different groups, because some content might resonate within a specific group but not outside the group. Yet since the vast majority of popular tweets do not travel between the activist and sceptic groups in our data, in this study we only focus on within-group virality: that is, for each tweet we only look at how virally it spreads in the group that it mainly spreads in. This is achieved by simply disregarding the follower network and retweets outside of the main-spreading group when inferring the virality of a tweet, in effect discarding both the successful and failed cascade events for users outside the group. 


To conduct the virality inference on our data, we built the follower network of the classified subset of users. Using the Twitter API, we first collected the Twitter followees of each user; and then we constructed the follower network with each user as a single node, and a link from each user to each of their followees. Following such process, we obtained a user follower network of 20,628 nodes (i.e. users) and 2,398,028 links, with which we were then able to infer the virality of every tweet in our data set.

\subsection{Viral themes among activists and sceptics}
\label{sec:viral}
The virality scores and the tweets we manually labelled in Section~\ref{sec:labeling} can be used to investigate which characteristics of a tweet make it viral. We explored this question by finding out
which characteristics of a tweet best predict its virality, in respectively the activist group and the sceptic group. More precisely, we fit a group lasso model \citep{yuan2006model} for each group, with the log-transformed virality score of a tweet as the response variable, and the tweet characteristics as explanatory variables. To better observe and control for author's effect on tweet virality, both in terms of the characteristics of the author and the structure of their follower network, we first selected a subset of the top 500 tweet set for this study, where each unique author has at least 3 tweets in the subset; the selected subset contains 261 tweets from the activist group and 128 tweets from the sceptic group. Then, we included among the explanatory variables a binary author indicator variable for each unique author in the selected subset. We then grouped all the explanatory variables so that all author indicator variables were in the same group, and each of the remaining variables was in a separate group.

A lasso model fits the data in a way similar to a linear regression model, yet additionally performs regularisation by removing redundant explanatory variables, so as to decrease the risk of overfitting an excessively complex model to noises in the sampled data, thereby overestimating the significance of effects, and reporting findings that will not replicate in the population \citep{babyak2004you, hawkins2004problem, mcneish2015using}. Consequently, this automatic variable selection process improves the parsimony and generalisability of the model, and the validity of its interpretations \citep{fariss2018enhancing, mcneish2015using}. Further, a group lasso model performs variable selection in a grouped manner, so that each group of variables is included or excluded as a whole. 

We selected with cross validation the most appropriate level of regularisation, with which the model had the best predictive performance on the validation set that was unseen during model training. Estimated coefficients were then interpreted in the same way as those from linear regression models with log-transformed outcomes. More specifically, a coefficient value of $x$ indicates that for every one-unit increase in the explanatory variable, the virality of the tweet is predicted to change by $(e^x-1)\cdot 100$\%. Here we present in Table~\ref{tab:lasso} the transformed coefficients, which can be directly interpreted as the predicted percentage change in tweet virality for every unit increase in the explanatory variables.

\begin{table}[!htb]
    \centering
    \small
    \begin{tabular}{lr|lr}
    \toprule
    \multicolumn{2}{c|}{Activist Group ($n$=261)} & \multicolumn{2}{c}{Sceptic Group ($n$=128)} \\
    \midrule
    Explanatory & Predicted Change & Explanatory & Predicted Change \\
    Variable & in Virality (\%) & Variable & in Virality (\%) \\
    \midrule
    Mocking & 0.0 & Mocking & 0.0 \\
    / & / & \textbf{Incivility} & 16.2 \\
    Call to Action & 0.0 & \textbf{Call to Action} & 14.5 \\
    \textbf{Ingroup Praise} & 0.9 & Ingroup Praise & 0.0 \\
    \textbf{Outgroup Criticism} & 1.5 & \textbf{Outgroup Criticism} & -7.9 \\
    \textbf{Science} & -2.7 & Science & 0.0\\
    \textbf{Movement} & 12.3 & / & / \\
    Solution & 0.0 & / & / \\
    / & / & Anti-international & 0.0 \\
    / & / & \textbf{Hypocrisy} & 10.1 \\
    / & / & Conspiracy & 0.0 \\
    \textbf{Hashtags} & 0.5 & \textbf{Hashtags} & 1.6 \\
    \textbf{Mentions} & 1.7 & \textbf{Mentions} & 1.1 \\
    \textbf{Authors} & -32.3$\sim$46.4 & \textbf{Authors} & -27.3$\sim$26.5\\
    \bottomrule
    \end{tabular}
    \caption{The predicted percentage change in tweet virality for every unit increase in the explanatory variables. $n$ indicates the number of samples used to fit the model in each group. The bold variables indicate those not excluded by the lasso model.}
    \label{tab:lasso}
\end{table}

We first looked at what characteristics of a tweet are related to its virality in the activist group. As shown in Table~\ref{tab:lasso}, the best model selected with the group lasso method excludes \textit{Mocking} and \textit{Solution}, indicating that whether a tweet in the activist group uses a mocking tone or includes a solution to climate change is a poor predictor of its virality. Among the retained features, the virality of a tweet seems to be mostly predicted by who the author of the tweet is, seen in the large effect estimates of the author indicator variables. The number of hashtags and mentions in a tweet are shown to be positively contributing to the virality of it, which resonates with previous work \citep{suh2010want}.

Among the tweet characteristics we labelled, \textit{Movement} seems to be the most important one in predicting the virality of a tweet in the activist group. The model indicates that if the tweet discusses the strength of a pro-climate movement, then its virality increases by 12.3\% in the activist group. Meanwhile, \textit{Outgroup Criticism} and \textit{Ingroup Praise} are also shown to be positively predicting the virality of a tweet, yet with a much weaker relationship. Interestingly, the occurrence of \textit{Science} seems to have a small negative effect on the virality of a tweet in the activist group. Controlling for other features, if the tweet invokes science as support, then its virality decreases by 2.7\%.

To more intuitively illustrate these effects, we present in Table~\ref{tab:pos-eg} several pairs of the most viral tweet and the least viral tweet posted by the same user from our data set. As shown, the most viral tweet by user A1 describes the strength of a pro-climate movement (thus labelled \textit{Movement}) and has a virality score of 0.155. On the other hand, A1's least viral tweet discusses the seriousness of the current air pollution situation with scientific evidence (mostly in the video attached) and calls for action to deal with it (thus labelled \textit{Science} and \textit{Call to Action}). It has a virality score of only 0.035. Meanwhile, the most-viral tweet by user A2, which praises Greta Thunberg and criticises her opponents (thus labelled \textit{Ingroup Praise} and \textit{Outgroup Criticism}), obtains a virality score of 0.125. A2's least viral tweet, similar to A1's, also uses scientific evidence to show the seriousness of climate change (thus labelled \textit{Science}), and only has a virality score of 0.035.

\begin{table}[!htb]
    \centering
    \footnotesize
    \begin{tabular}{p{0.05\columnwidth}|p{0.43\columnwidth}|p{0.43\columnwidth}}
    \toprule
    User & Most viral tweet & Least viral tweet \\
    \midrule
    A1 & \textbf{Virality score: 0.155} & \textbf{Virality score: 0.035} \\
    & \textbf{[Movement]} & \textbf{[Science, Call to Action]} \\
    & LOOK AT ALL THOSE PEOPLE marching in Alberta to demand \#climateaction! Even here people want action on the \#ClimateEmergency.
    \newline
    Denying our situation doesn't help. \#ActOnClimate
    \newline
    \#ClimateStrike \#FridaysforFutures \#cdnpoli \#climate \#energy \#elxn43 @GretaThunberg. Via @vineshpratap
    \newline 
    [video] & 
    Air pollution is now more deadly than war, smoking and TB. It kills 7 million people every year.
    \newline
    We have solutions to keep our communities safe and deal with the \#climate crisis. Let's implement them. \#GreenNewDeal
    \newline
    \#AirPollution \#ClimateChange \#energy \#tech \#PanelsNotPipelines 
    \newline
    [video] \\
    \midrule
    A2* & \textbf{Virality score: 0.125} & \textbf{Virality score: 0.035} \\
    & \textbf{[Ingroup Praise, Outgroup Criticism]} & \textbf{[Science]} \\
    &
    Attacks on Thunberg is motivated by one thing. She is intelligent, eloquent, compassionate, and young. She has scared some hateful and reactionary so-called `grown ups',
    \newline
    \#ActOnClimate \#ClimateCrisis
    \newline
    [link] & 
    Climate change has rapidly and dramatically affected the arctic region. Even just ten years ago it was impossible for container ships to go through the Northern Sea Route.
    \newline
    [link] \\
    \bottomrule
    \end{tabular}
    \caption{Pairs of the most viral tweet and the least viral tweet posted by the same user, within the selected tweets in the activist group. The virality score and labels of each tweet are shown along with the text. Tweets from non-public figures, marked with an *, are paraphrased while retaining the same features to protect users' privacy.}
    \label{tab:pos-eg}
\end{table}

With respect to the sceptic group, the best model selected with the group lasso excludes \textit{Ingroup Praise}, \textit{Science}, \textit{Anti-international}, and \textit{Conspiracy}, suggesting that such themes are not useful in predicting the virality of a tweet in the sceptic group. Similarly as in the activist group, the identity of the tweet author is the strongest predictor of the virality of a tweet, and the number of mentions and hashtags have positive effects. 

Among the labelled characteristics, \textit{Incivility} and \textit{Call to Action} seem to be the most important positive predictors of tweet virality in the sceptic group, while \textit{Hypocrisy} also has a positive effect. Specifically, if the tweet contains uncivil language, then its virality increases by 16.2\% in the sceptic group. If it contains calls to action, its virality increases by 14.5\%. If it contains hypocrisy claims, its virality increases by 10.1\%. Intriguingly, despite the overwhelming number of these tweets, the \textit{Outgroup Criticism} feature predicts decreased tweet virality in the sceptic group. If a tweet makes an outgroup criticism, then its predict virality decreases by 7.9\%. 

In Table~\ref{tab:neg-eg}, we show the most viral and least viral tweet pairs within the selected tweets in the sceptic group. Specifically, the most viral tweets by user S1, S2 and S3, all with virality scores over 0.11, respectively involve uncivil language, calls to action, and hypocrisy claims, which are the three most significant characteristics we find to be positively predicting virality in the sceptic group. Meanwhile, the least viral tweets from the three users involve characteristics that are shown to have negative or no effect on tweet virality, such as \textit{Mocking}, \textit{Conspiracy} and \textit{Outgroup Criticism}. We can also observe that \textit{Outgroup Criticism} co-occurs with viral themes (e.g. \textit{Hypocrisy}) in viral tweets, but occurs by itself in less viral ones -- which in part explains the predicted negative effect of \textit{Outgroup Criticism} on virality.

\begin{table}[!htb]
    \centering
    \footnotesize
    \begin{tabular}{p{0.05\columnwidth}|p{0.43\columnwidth}|p{0.43\columnwidth}}
    \toprule
    User & Most viral tweet & Least viral tweet \\
    \midrule
    S1* & \textbf{Virality score: 0.190} & \textbf{Virality score: 0.037} \\
    & \textbf{[Incivility]} & \textbf{[Mocking]} \\
    & 
    I learned basic skills like sewing, cooking, woodwork, automobiles, and metalwork in high school home economics and shop classes :)
    \newline
    Kids are now taught BS Socialism and Fake Climate Change... :/
    \newline
    KAG &
    Thanks to California banning Plastic Straws... The Climate is becoming much better... It's only 45 today when it would have been 46... It’s working... yay... \\
    \midrule
    S2* & \textbf{Virality score: 0.118} & \textbf{Virality score: 0.044} \\
    & \textbf{[Call to Action]} & \textbf{[Conspiracy, Outgroup Criticism]} \\
    & 
    Just one year ago, no political party would say:
    \newline
    - Climate crisis is a hoax
    \newline
    - Immigration is too high
    \newline
    - Corporate welfare and supply management must be elimated
    \newline
    - The budget can be balanced within two years
    \newline
    PPC is changing the conversation and bringing change along with it.
    \newline
    \#VotePPC & 
    The network of globalist elites are all supporting each other. They promote fake climate change to scare people into submission. They are liars. \\
    \midrule
    S3* & \textbf{Virality score: 0.113} & \textbf{Virality score: 0.072} \\
    & \textbf{[Hypocrisy, Outgroup Criticism]} & \textbf{[Outgroup Criticism]} \\
    &  
    Greens Sarah Hanson-Young: The government needs to declare a climate emergency!
    \newline
    Also Sarah Hanson-Young: I have taken 58 flights this year with taxpayer money.
    \newline
    Do as I say, not do as I do.
    \newline
    [image] & 
    When will the Victoria police bill all these climate change protestors?
    \newline
    [link] \\
    \bottomrule
    \end{tabular}
    \caption{Pairs of the most viral tweet and the least viral tweet posted by the same user, within the selected tweets in the sceptic group. The virality score and labels of each tweet are shown along with the text. Tweets from non-public figures, marked with an *, are paraphrased while retaining the same features to protect users' privacy.}
    \label{tab:neg-eg}
\end{table}

\section{Discussion}
Our findings on virality-related characteristics first complement our empirical observations of common climate discussion themes (Section~\ref{sec:labeling}) in confirming a potential shifting trend of focus in the Twitter climate debate. Compared with \posscite{pearce2014climate} analysis of Twitter discussions around the 2013 IPCC report, which found “science” to be one of the most prominent themes, our results show that the discussion of climate change science, although still an active topic, is not among the most popular themes (Figure~\ref{fig:popwords}), and is even among the least viral themes (Table~\ref{tab:lasso}) in both groups. The viral themes we found, including \textit{Movement} among activists and \textit{Incivility}, \textit{Hypocrisy} among sceptics, further suggest that the core of Twitter climate discourse might have switched from the existence of climate change to climate activism -- either emphasizing the climate movements from the activist side, or attacking climate activism from the sceptic side.

On the other hand, our results also show the difference between popular themes and viral themes. Specifically, the use of uncivil language, although not a dominating theme in the sceptic group (Figure~\ref{fig:var}), is the most important characteristic that predicts tweet virality among the sceptics (Table~\ref{tab:lasso}). In the meantime, outgroup criticism themes increase tweet virality in the activist group where they are less prevalent, and decrease tweet virality in the sceptic group where they are more prevalent. Apart from the theme co-occurrence issue that we discussed at the end of Section~\ref{sec:viral}, such discrepancy might be related to the different types of entities under criticism in the two groups. In the activist group, criticism is mostly directed at the governments, which is democratically understood to be a socially acceptable target of dissatisfaction. In the sceptic group, however, criticism is usually directed at certain individuals and accompanied by mocking, incivility, or hypocrisy claims, thus potentially only attracting support from a limited subgroup of people. Such phenomenon also aligns with the sensation seeking literature \citep{bench2013function, zuckerman2010sensation} in suggesting that people tend to lose interest in patterns under repeated exposure, and are more easily elicited by novel stimuli.

Finally, we consider together the most important predictor of tweet virality in each group: \textit{Movement} among activists and \textit{Incivility} among sceptics. Clearly, they first resonate with our common theme analysis (Section~\ref{sec:labeling}) in showing the heterogeneity of discourse in the two groups, and potentially hinting at different types of links that tie the community together. More specifically, Twitter climate activists nowadays might be most effectively united through the discussion of climate movements, while Twitter climate sceptics most likely bond over shared hostility toward climate activism.

Despite their evident difference, the \textit{Movement} and \textit{Incivility} themes both seem to serve the function of enhancing emotional connections within the group while rejecting potential involvement from outside the group. Specifically, promoting pro-climate movements likely increases enthusiasm from climate activists while escalating the resistance of climate sceptics toward the issue. Meanwhile, incivility in comments against climate activism likely amplifies resonance among sceptics, yet elicit contempt and animosity from activists. Following \posscite{dahlgren2021critical} claim that people are constantly exposed to information from the outgroup, which originally served as evidence against polarisation, our results instead show that this exposure probably exacerbates polarisation, echoing \posscite{bail2018exposure} similar finding from a field experiment. In this sense our findings reveal, from an information spreading perspective, how the echo chambers of climate discussions potentially get further consolidated and separated. 


\section{Conclusion}
In this paper we set out to study climate change discussions on Twitter through the lens of information spreading. Our work first presents an up-to-date picture of popular climate communication themes on Twitter both within and across the activist and sceptic groups. We show that climate activists and climate sceptics generally communicate within their own groups in disparate styles, and we additionally find a virtual absence of information sharing across the groups. These results corroborate prior findings that show evidence of echo chambers in climate communication on Twitter \citep{williams2015network}. 

More importantly, we make a distinct contribution by examining the tweet characteristics that predict viral spreading within the two groups. First, we find that the virality-predicting themes showcase interesting matches and mismatches with the popular themes. Further interpreting the strongest predictors of viral spreading -- \textit{Movement} among activists and \textit{Incivility} among sceptics -- we argue that while these themes reflect different types of bonds that tie the community together, they both tend to enhance ingroup connections while repulsing outgroup engagement. This finding has implications in the broader context of climate change politics and communication, in that it reveals the potential for viral spreading to exacerbate polarisation in the climate debate on Twitter.


\section*{Acknowledgements}
All the authors acknowledge funding from the Academy of Finland, project ECANET, number 320781. In addition, TC acknowledges Academy of Finland, project ECANET, number 320780. The authors also acknowledge the computational resources provided by the Aalto Science-IT project.

\small\singlespacing\RaggedRight
\Urlmuskip=0mu plus 1mu\relax
\bibliography{virality}

\end{document}